\documentclass[doublecol]{epl2}
\usepackage{graphicx} 
\usepackage{amsmath}
\usepackage{amsfonts}
\usepackage{amssymb}

\usepackage{color}
\definecolor{myred}{rgb}{1,0,0}
\newcommand{\ket}[1]{|#1\rangle}
\newcommand{\bra}[1]{\langle #1|}
\newcommand{\Tr}{\text{Tr}}

\title{Experimentally testable geometric phase of sequences of 
Everett's relative quantum states}
\author{Erik Sj\"oqvist}
\institute{Department of Quantum Chemistry, Uppsala University, 
Box 518, Se-751 20 Uppsala, Sweden}
\abstract{Everett's concept of relative state is used to introduce a 
geometric phase that depends nontrivially on entanglement in a pure 
quantum state. We show that this phase can be measured in multiparticle
interferometry. A correlation-dependent generalization of the relative
state geometric phase to mixed quantum states is outlined.}
\begin{document}
\maketitle
\section{Introduction}
Pancharatnam's geometric phase \cite{pancharatnam56} is a property of
a discrete set of polarization states obtained by sending light beams
through a polarization analyzer. This geometric phase has a natural
counterpart in quantum mechanics, namely the phase that arises when a
quantal system is exposed to a sequence of filtering measurements
\cite{samuel88,benedict89,cassinelli94}. Geometric phases associated 
with discrete sequences of quantum states have been considered in
various contexts such as quantum Zeno effect \cite{facchi99}, quantum
jumps \cite{carollo03}, and weak measurements \cite{sjoqvist06b}

Here, we wish to consider another physical context in which discrete
geometric phases may occur. We propose a notion of geometric phase
associated with sequences of Everett's relative quantum states
\cite{everett57}. The essential property of this geometric phase is
that it relates directly to correlations and entanglement; as such it
connects to earlier studies of geometric phases for time evolving
entangled systems
\cite{sjoqvist00a,milman03,tong03,wang05,basu06,williamson07,chen07}. 
We wish to examine the geometric phase of relative quantum states 
and demonstrate how it can be unveiled experimentally.

\section{Geometric phase of relative states.}
For some composite quantum system, consider a state $\Psi$
represented by the normalized vector $\ket{\Psi} \in \mathcal{H}$. 
Assume a partitioning into subsystems defined by the
tensor product decomposition $\mathcal{H} = \mathcal{H}_1 \otimes
\mathcal{H}_2$. Let $P(\mathcal{H})$, $P(\mathcal{H}_1)$, and 
$P(\mathcal{H}_2)$ be the corresponding ray (state) spaces.

Now, suppose a projection measurement on the first subsystem 
yields the state $\phi \in P(\mathcal{H}_1)$. Given this outcome, 
the post-measurement state can be calculated by applying the 
projection operator $\ket{\phi} \bra{\phi} \otimes
\hat{1}$ to $\ket{\Psi}$. This induces the transformation 
$\ket{\Psi} \rightarrow \ket{\phi} \ket{\Psi (\phi)}$ with 
$\ket{\Psi (\phi)}$ being the partial scalar product 
$\bra{\phi} \Psi \rangle \in \mathcal{H}_2$. Here, $\ket{\Psi (\phi)}$ 
defines Everett's relative state \cite{everett57} 
in the second subsystem given the outcome $\phi$ in the first 
subsystem. If $\phi$ is normalized, the squared norm of 
$\Psi (\phi)$ is the marginal probability to obtain $\phi$ given 
$\Psi$, i.e., $\parallel \! \! \ket{\Psi (\phi)} \! \! \parallel^2 = 
\bra{\phi} \rho_1 \ket{\phi}$ with $\rho_1 = \Tr_2 \ket{\Psi} 
\bra{\Psi}$ the marginal state on $\mathcal{H}_1$ ($\Tr_2$ is 
partial trace over system $2$). The relative state may be formulated
in a compact way in terms of the antilinear map $\mathcal{L}_{\Psi}:
\mathcal{H}_1 \rightarrow \mathcal{H}_2,\ \ket{\phi} \mapsto 
\bra{\phi} \Psi \rangle$ \cite{kurucz01}. 

Pancharatnam's \cite{pancharatnam56} notion of ``in-phase'' is the
underlying principle of the geometric phase. For two nonorthogonal
states $\phi_a$ and $\phi_b$ in $P(\mathcal{H}_1)$, the vector
representatives $\ket{\phi_a}$ and $e^{if}\ket{\phi_b}$ in
$\mathcal{H}_1$ are in-phase if they produce a maximum in intensity
when superposed. This corresponds to the condition $e^{-if}
\bra{\phi_b} \phi_a \rangle >0$. Solving for $f$ yields the
Panharatnam phase $f=\arg \bra{\phi_b} \phi_a \rangle$. Now,
the superposition $\ket{\phi_a}+e^{if}\ket{\phi_b}$ is mapped by
$\mathcal{L}_{\Psi}$ to $\ket{\Psi (\phi_a)} + e^{-if}
\ket{\Psi (\phi_b)}$. Provided $\Psi (\phi_a)$ and $\Psi (\phi_b)$ are
nonorthogonal, the in-phase condition yields the Pancharatnam phase
\begin{eqnarray}
f = \arg \bra{\Psi (\phi_a)} \Psi (\phi_b) 
\rangle = \arg \bra{\phi_b} \rho_1 \ket{\phi_a} \rangle 
\label{eq:panchrelph}
\end{eqnarray}
over 
the relative states $\Psi (\phi_a)$ and $\Psi (\phi_b)$. 
Note the antilinearity-induced interchange $\phi_a 
\leftrightarrow \phi_b$ in Eq. (\ref{eq:panchrelph}). 

We are now prepared to introduce the geometric phase for a sequence of
relative states for a given bipartite state $\Psi \in P(\mathcal{H}_1
\otimes \mathcal{H}_2)$. Let $L: \phi_1,\ldots,\phi_N$ be an ordered
sequence of states in $P(\mathcal{H}_1)$ and let $\ket{\phi_1}, \ldots
, \ket{\phi_N} \in \mathcal{H}_1 - \{ 0 \}$ be vectors over $L$. 
$L$ defines the sequence $\Psi (L) : \Psi (\phi_1), \ldots , 
\Psi (\phi_N)$ of relative states in $P(\mathcal{H}_2)$. Assume 
all adjacent pairs in $\Psi (L)$ are nonorthogonal. We may assign 
a geometric phase $\gamma [\Psi (L)]$ to the sequence $\Psi (L)$ of 
relative states by using the Bargmann prescription \cite{mukunda93}
and Eq. (\ref{eq:panchrelph}), yielding
\begin{eqnarray} 
\gamma [\Psi (L)] & = & 
\arg \big( \langle \Psi (\phi_1) \ket{\Psi (\phi_N)} 
\nonumber \\ 
 & & \times \langle \Psi (\phi_N) \ket{\Psi (\phi_{N-1})} \cdots 
\langle \Psi (\phi_2) \ket{\Psi (\phi_1)} \big) 
\nonumber \\ 
 & = & - \arg \big( \bra{\phi_1} \rho_1 
\ket{\phi_N} 
\nonumber \\ 
 & & \times \bra{\phi_N} \rho_1 \ket{\phi_{N-1}} \cdots 
\bra{\phi_2} \rho_1 \ket{\phi_1} \big) .     
\label{eq:relgp1}
\end{eqnarray}
Similarly, we can define the geometric phase associated with 
$L$ as   
\begin{eqnarray} 
\gamma [L] = \arg \big( \bra{\phi_1} \phi_N \rangle 
\bra{\phi_N} \phi_{N-1} \rangle \cdots \bra{\phi_2} \phi_1 \rangle  
\big)   
\label{eq:gp}   
\end{eqnarray}
provided the nonorthogonality between adjacent pairs is satisfied 
also along $L$. The antilinearity-induced interchange $\phi_a 
\leftrightarrow \phi_b$ mentioned above, results
in an effective reversal of the path in $P(\mathcal{H}_2)$, which is
the origin of the sign difference in the right-hand side of
Eqs. (\ref{eq:gp}) and (\ref{eq:relgp1}). $\gamma [\Psi (L)]$ and 
$\gamma [L]$ are invariant under the local gauge 
transformation $\ket{\phi_j} \rightarrow c_j \ket{\phi_j}$, 
$c_j \in \mathbb{C} - \{ 0 \}$ for $j=1,\ldots,N$.

Let us now see how the geometric phase of $\Psi (L)$ depends on 
entanglement. For a product state (no entanglement), $\rho_1$ is a
pure projector, which applied to Eq. (\ref{eq:relgp1}) implies that
$\gamma [\Psi (L)]$ vanishes. Thus, the phase $\gamma [\Psi (L)]$
is zero if $\Psi$ is a product state.
For a maximally entangled state
$\Psi$ (assuming $K =\dim \mathcal{H}_1 \leq \dim \mathcal{H}_2$
finite), $\rho_1 = \frac{1}{K} \hat{1}$ and we obtain 
$\gamma [\Psi (L)] = -\gamma [L]$. This relation between 
$\gamma [\Psi (L)]$ and $\gamma [L]$ is a consequence of the 
antiunitary (``time-reversal'') nature of $\sqrt{K} \mathcal{L}_{\Psi}$ 
for maximally entangled $\Psi$ \cite{kurucz01}. 

We may put $\gamma [\Psi (L)]$ on integral form by using the concept
of null phase curves \cite{rabei99}. These are defined as curves that
have vanishing geometric phase for any portion of them. A free
geodesic is always a null phase curve, but the converse is not
true, in general. Free geodesics and null phase curves fully coincide
only in the two-dimensional (qubit) case. Now, let the relative states
$\Psi (\phi_j)$, $j=1,\ldots,N$ be connected by null phase curves
forming a continuous path $\widetilde{L}$ path in ray space. Then,
following Ref.
\cite{rabei99}, we may write  
\begin{eqnarray}
\gamma [C] = \oint_{\widetilde{L}} A
\end{eqnarray}
with connection one-form 
\begin{eqnarray} 
A = {\textrm{Im}} \frac{\bra{\phi} \rho_1 \ket{d\phi}}
{\bra{\phi} \rho_1 \ket{\phi}} .  
\end{eqnarray}

\begin{figure}[ht!] 
\begin{center} 
\includegraphics[width=8.5 cm]{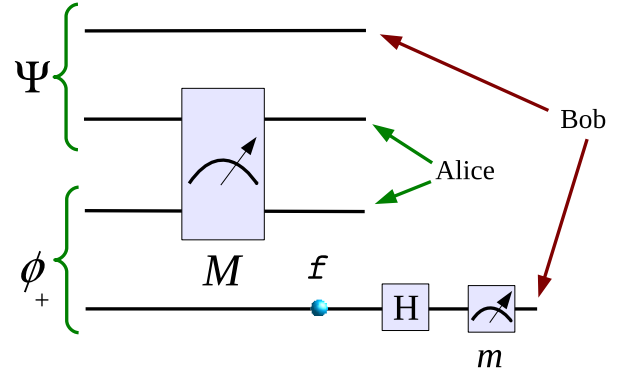} 
\end{center} 
\caption{\label{fig:swapping} Physical realization of geometric 
phase $\gamma [\Psi (L)]$ relative $L:\phi_1,\ldots,\phi_N$. Alice and
Bob share $\Psi$ and a Bell state $\ket{\Phi_+} = \frac{1}{\sqrt{2}}
(\ket{00} + \ket{11})$. Alice performs a measurement $M$ and
post-selects the entangled state $\ket{\phi_j} \ket{0} + e^{if_j}
\ket{\phi_{j+1 \! \! \mod N}}
\ket{1}$, where $\phi_j$ and $\phi_{j+1 \! \! \mod N}$ constitute 
two nearby states along $L$ and $f_j$ maximizes the intensity 
measured by Bob in step $j-1$ with $f_1 = 0$. Alice communicates 
to Bob which members of the ensemble should be kept. Bob performs 
an interference experiment on the resulting subensemble and finds 
$f_{j+1}$ by maximizing the intensity by varying $f$. Iteration 
yields $f_{N+1}=\gamma [\Psi (L)]$.}
\end{figure} 

\section{Physical realization} 
We demonstrate how the geometric phase $\gamma [\Psi (L)]$ can be
measured (Fig. \ref{fig:swapping} shows a circuit version of the
setup). The key step is to implement the Pancharatnam phase of
adjacent pairs of relative states. This can be done by consuming
entangled pairs of extra ancillary qubits.

Prepare the pure state $\ket{\Xi} =
\ket{\Psi} \ket{\Phi_+}$, where $\ket{\Phi_+} =
\frac{1}{\sqrt{2}} (\ket{00}+\ket{11})$ is a Bell state of two 
ancilla qubits. Alice (Bob) possesses the first (second) half of 
$\Psi$ and one of the ancilla qubits. Alice performs a measurement 
$M$ that realizes the projection operator     
\begin{eqnarray} 
 & \Pi_{j+1,j} = 
\frac{1}{2} \left( \ket{\phi_j} \ket{0} + 
e^{if_j} \ket{\phi_{j+1 \! \! \! \mod N}} \ket{1} \right)
\nonumber \\ 
 & \times \left( \bra{\phi_j} \bra{0} + 
e^{-if_j} \bra{\phi_{j+1 \! \! \! \mod N}} \bra{1} \right) , 
\end{eqnarray}  
which swaps the entanglement into the state 
\begin{eqnarray} 
 & \Pi_{j+1,j} \otimes \hat{1} \ket{\Xi} =  
\frac{1}{2} \left( \ket{\phi_j} \ket{0} + 
e^{if_j} \ket{\phi_{j+1 \! \! \! \mod N}} \ket{1} \right) 
\nonumber \\ 
 & \times 
\left( \ket{\Psi(\phi_j)} \ket{0} + 
e^{-if_j} \ket{\Psi(\phi_{j+1 \! \! \! \mod N})} \ket{1} \right) .
\end{eqnarray}
Alice and Bob can do independent interference experiments on the 
post-selected ensemble described by $\Pi_{j+1,j} \otimes \hat{1} 
\ket{\Xi}$. Bob's experiment relates to  
$\gamma [\Psi (L)]$ as follows. He first applies a phase 
shift $f:\ket{x} \rightarrow e^{ixf} \ket{x}$, $x=0,1$, followed by a 
Hadamard ${\textrm{H}}: \ket{x} \rightarrow \frac{1}{\sqrt{2}} 
(\ket{x} + (-1)^x \ket{x\oplus 1})$, and finally performs a measurement 
$m$ of the $0$ state, say. This intensity is maximal 
for $f = \arg \bra{\Psi (\phi_{j+1 \! \! \mod N})} \Psi (\phi_j) \rangle + 
f_j \equiv f_{j+1}$. Iteration with $f_1 = 0$ yields $f_{N+1} = 
\gamma [\Psi (L)]$. 

We note that Alice and Bob must communicate classically in order to
measure $\gamma [\Psi (L)]$. In this sense, $\gamma [\Psi (L)]$ is a
nonlocal quantity that reflects entanglement in pure quantum
states. On the other hand, the marginal subsystem geometric phase is a
local quantity since no classical communication is needed to measure
it (see, e.g., Fig. 1 of Ref. \cite{tong03}).

Furthermore, note that one may replace the ancillary Bell state
$\Phi_+$ by a nonmaximally entangled state $\ket{\Phi} = \sqrt{a}
\ket{00} + \sqrt{1-a} \ket{11}$, $a\in [0,1]$, at the expense of 
reducing the interference visibility by a factor $2\sqrt{a(1-a)}$. 
This factor varies from $0$ (product states $a=0,1$) to the optimal 
value $1$ (Bell state $a=\frac{1}{2}$).

\begin{figure}[ht!] 
\begin{center} 
\includegraphics[width=8.5 cm]{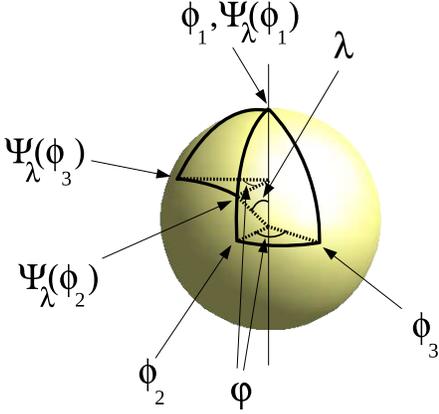} 
\end{center} 
\caption{\label{fig:blochsphere} Sequence of qubit states 
$L:\phi_1,\phi_2,\phi_3$ and their relative states $\Psi_{\lambda}(L):
\Psi_{\lambda} (\phi_1),\Psi_{\lambda} (\phi_2),\Psi_{\lambda}(\phi_3)$ 
for the entangled two-qubit state $\Psi_{\lambda}$ in
Eq. (\ref{eq:2q}). The two discrete sets of states are joined by null
phase curves, which coincide with geodesics (great circle segments) on
the Bloch sphere. The sequences $L$ and $\Psi_{\lambda}(L)$ enclose
the (signed) solid angles $\varphi$ and $-\varphi + 2\arctan \left(
\cos \lambda \tan \frac{\varphi}{2} \right)$, 
respectively.}
\end{figure} 

\section{Examples} 
First, consider a two-qubit state 
\begin{eqnarray}
\ket{\Psi_{\lambda}} = 
\cos \frac{\lambda}{2} \ket{00} + \sin \frac{\lambda}{2} \ket{11} , \ 
\lambda \in [0,\pi] 
\label{eq:2q}
\end{eqnarray} 
with maximal entanglement for $\lambda = \frac{\pi}{2}$ and product
state for $\lambda = 0$ or $\pi$. This state may be prepared
experimentally in the polarization of two photons using spontaneous
down-conversion technique \cite{white99}.  Consider the sequence $L:
\phi_1,\phi_2,\phi_3$ connected by null phase curves \cite{rabei99}
(here parts of great circles) on the Bloch sphere, as shown in
Fig. \ref{fig:blochsphere}. Explicitly, $\phi_1,\phi_2,\phi_3$ are
represented by the vectors $\ket{0} , \frac{1}{\sqrt{2}}
\big( \ket{0} + \ket{1} \big) , \frac{1}{\sqrt{2}} \big( \ket{0} + 
e^{-i\varphi}\ket{1} \big) \in \mathcal{H}_1$, respectively, with
$0\leq \varphi <\pi$. The geometric phase becomes $\gamma [L] =
-\frac{1}{2} \varphi$. The vectors $\cos \frac{\lambda}{2} \ket{0} ,
\frac{1}{\sqrt{2}} \big( \cos \frac{\lambda}{2}
\ket{0} + \sin \frac{\lambda}{2} 
\ket{1} \big) , \frac{1}{\sqrt{2}} \big( \cos \frac{\lambda}{2} \ket{0} + 
e^{i\varphi} \sin \frac{\lambda}{2} \ket{1} \big) \in \mathcal{H}_2$ 
represent the members of the sequence $\Psi_{\lambda} (L) : 
\Psi_{\lambda} (\phi_1), \Psi_{\lambda} (\phi_2),\Psi_{\lambda} (\phi_3)$ 
relative $L$ (again see Fig. \ref{fig:blochsphere}). The corresponding 
geometric phase reads  
\begin{eqnarray}
\gamma [\Psi_{\lambda} (L)] = \frac{\varphi}{2} - 
\arctan \left( \cos \lambda \tan \frac{\varphi}{2} \right) ,  
\end{eqnarray}
which varies from $0$ to $\frac{\varphi}{2} = -\gamma (L)$ when
$\lambda$ increases from $0$ (product state) and $\frac{\pi}{2}$
(maximally entangled state). Note that $\gamma [\Psi_{\lambda} (L)]$
is undefined for this specific $L$ if $\lambda = \pi$.

\begin{figure}[ht!] 
\begin{center} 
\includegraphics[width=8.5 cm]{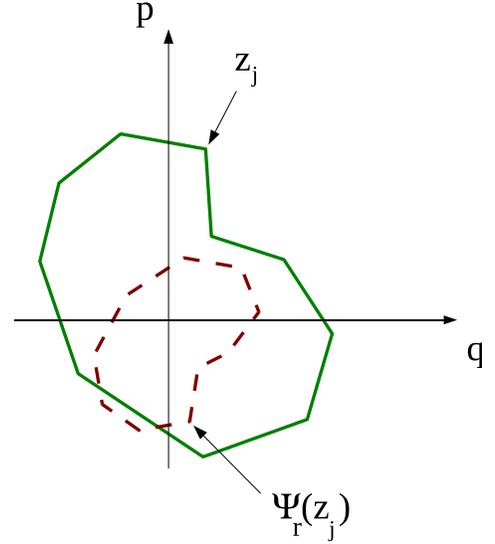} 
\end{center} 
\caption{\label{fig:gaussian} Sequence of coherent states 
$L: z_1,\ldots,z_N$ (solid line) and their relative states $\Psi_r
(L):\Psi_r (z_1),\ldots,\Psi_r (z_N)$ (dotted line) for the entangled
two-mode squeezed state $\Psi_r$ in Eq. (\ref{eq:r}). The two discrete
sets of states are joined by null phase curves, which are straight
lines in the complex plane. Each coherent state $z_j$ defines a point
$(q_j,p_j)$ in phase space via the relation $z_j=\frac{1}{\sqrt{2}}
(q_j+ip_j)$. We find that $\Psi_r (z_j) =
\tanh(r) z_j^{\ast}$, i.e., scaled coherent states mirror reflected through 
the $q$-axis.  The sequences $L$ and $\Psi_r (L)$ enclose the (signed) 
phase space areas $\oint_{\widetilde{L}} pdq$ and $-\tanh^2 (r) 
\oint_{\widetilde{L}} pdq$, respectively, where $\widetilde{L}$ is 
the geodesic polygon defined by $L$.}
\end{figure} 

Our second example concerns two-mode squeezed states \cite{giedke03}: 
\begin{eqnarray} 
\ket{\Psi_r} = \frac{1}{\cosh (r)} \sum_{n=0}^{\infty} \tanh^n (r) 
\ket{nn}
\label{eq:r}
\end{eqnarray} 
with $r\geq 0$ the squeezing parameter, which determines the amount
of entanglement in $\Psi_r$ ($r=0$ corresponds to no
entanglement and $\Psi_r$ tends to a maximally entangled state when
$r\rightarrow \infty$). This class of states may be prepared 
experimentally by sending a 
down-converted photon-pair through optical fibers \cite{kimble87}. 
Consider a sequence $L : z_1,\ldots,z_N$ of coherent states
of the first oscillator mode, where each $z_j$ defines a point
$(q_j,p_j)$ in phase space. Explicitly, $z_j = \frac{1}{\sqrt{2}}
(q_j+ip_j)$. The null phase curves that join adjacent $z_j$'s become
straight lines, defining an $N$-vertex polygon $\widetilde{L}$ in
phase space, as shown in Fig. \ref{fig:gaussian}. Note that these null
phase curves correspond to constrained geodesics in the infinite
dimensional Hilbert space of the oscillator mode. The geometric phase
$\gamma [L]$ is the area $\oint_{\widetilde{L}} pdq$ enclosed by
$\widetilde{L}$. The corresponding sequence $\Psi_r (L)$ of relative
states of the second oscillator mode are coherent states $\Psi_r (z_j) =
\tanh(r) z_j^{\ast}$, related to $z_j$ by a scale factor $\tanh(r)$
and a mirror reflection through the $q$-axis (see Fig. \ref{fig:gaussian}). 
It follows that
\begin{eqnarray}
\gamma [\Psi_r (L)] = - \tanh^2 (r) \oint_{\widetilde{L}} pdq , 
\end{eqnarray}  
which interpolates between $0$ for product state $r=0$ and 
$-\gamma [L]$ when $r\rightarrow \infty$.  

\section{Mixed state case} 
The above analysis can be extended to mixed states by using the
Uhlmann holonomy \cite{uhlmann86}. Consider a mixed state represented
by the density operator $\varrho$ that acts on $\mathcal{H}_1 \otimes
\mathcal{H}_2$. This $\varrho$ is a product state if $\varrho = \Tr_2
\varrho \otimes \Tr_1 \varrho$; $\varrho$ is separable (classically
correlated) if it can be written as a convex sum of product states;
$\varrho$ is nonseparable (entangled) otherwise. The members of the
ordered sequence $\varrho (L) : \varrho (\phi_1) ,\ldots, \varrho
(\phi_N)$ relative $L : \phi_1 , \ldots ,\phi_N \in P(\mathcal{H}_1)$
are defined as $\varrho (\phi_j) = \Tr_1 \big[ \ket{\phi_j}
\bra{\phi_j}
\otimes \hat{1} \varrho \big] = \bra{\phi_j} \varrho \ket{\phi_j}$.  
For simplicity, suppose all $\varrho (\phi_j)$ are faithful, i.e., 
full rank. The Uhlmann holonomy for $\varrho (L)$ reads 
\begin{eqnarray} 
 & U_{\varrho (L)} = \left( \sqrt{\varrho (\phi_N)} \varrho (\phi_{N-1}) 
\sqrt{\varrho (\phi_N)} \right)^{-1/2} 
\nonumber \\ 
 & \times \sqrt{\varrho (\phi_N)} \sqrt{\varrho (\phi_{N-1})} \cdots 
\left( \sqrt{\varrho (\phi_2)} \varrho (\phi_1) 
\sqrt{\varrho (\phi_2)} \right)^{-1/2} 
\nonumber \\ 
 & \times \sqrt{\varrho (\phi_2)} 
\sqrt{\varrho (\phi_1)} . 
\end{eqnarray}
This quantity becomes trivially equal to the identity for product
states, but is in general nontrivial for separable and nonseparable
states. Thus, $U_{\varrho (L)}$ depends on both classical and quantum
correlations in the state; as such $U_{\varrho (L)}$ should be
regarded a correlation-dependent rather than entanglement-dependent
quantity of mixed bipartite states. It can in principle be measured
experimentally (see Ref. \cite{aberg07} for an explicit
interferometric scheme to observe the Uhlmann holonomy for an
arbitrary discrete sequence of faithful density operators).

\section{Conclusions} 
Geometric phases for discrete sequences of quantum states have been
considered in the past. Here, we have introduced such a concept for
sequences of Everett's relative states with respect to a bipartite 
tensor product decomposition of a pure quantum state. This phase depends 
on the amount of entanglement in the state, in particular it vanishes 
for product states. We have demonstrated that the geometric phase of
sequences of relative quantum states can be tested in multiparticle
interferometry, by means of local projective measurements and
classical communication assisted post-selection. We have argued that a
mixed state generalization based on the Uhlmann holonomy leads to a
multidimensional holonomy that depends on both classical and quantum
correlations in a noisy bipartite system.
\acknowledgments
Financial support from the Swedish Research Council is 
acknowledged. 

\end{document}